 \documentclass[preprint,12pt]{elsarticle}

\usepackage{amssymb}
\usepackage{lineno}
\journal{arxiv.org}
\biboptions{sort&compress} 
\usepackage[dvipsnames]{xcolor} 

\begin{document}

\begin{frontmatter}


\author[inst1,inst2]{H. D. Zhang\corref{cor1}}
\ead{haozhang@liverpool.ac.uk}
\cortext[cor1]{}
\affiliation[inst1]{organization={University of Liverpool},
            city={Liverpool},
            postcode={L69 7ZX}, 
            country={UK}}
\affiliation[inst2]{organization={Cockcroft Institute},
            city={Warrington},
            postcode={WA4 4AD}, 
            country={UK}}
\affiliation[inst3]{organization={CERN},
            city={Geneva},
            postcode={1211}, 
            country={Switzerland}}

\title{Characterization of a supersonic gas jet for charged particle beam profile monitor}




\author[inst1,inst2]{A. Salehilashkajani}
\author[inst1,inst2,inst3]{O. Sedlacek}
\author[inst1,inst2]{C. P. Welsch}

\begin{abstract}
In present work, we report an experimental method to measure the density and distribution of a supersonic gas jet used for charged particle beam profile monitoring. The density of the gas jet used in this monitor was in the range of $10^{14}$ - $10^{17}$  molecules/$m^3$. 
\end{abstract}



\begin{keyword}
supersonic gas jet \sep density measurement \sep non-invasive \sep beam profile monitor
\end{keyword}

\end{frontmatter}


\section{Introduction}
\label{sec:introduction}
Noninvasive measurement methods are preferred for modern accelerators to characterize the beam parameters. Ionization profile monitors (IPMs) \cite{AnneIPM1993,MiessnerIPM2010,BartkoskiIPM2014} and beam induced fluorescent monitors (BIFs) \cite{VariolaBIF2007,TsangBIF2013,ForckBIFDIPAC2003,ForckBIFIPAC2010,BeckerBIFDIPAC2007} are widely used as non-invasive beam profile monitors in many accelerators. In such monitors, the particle beams interact with the residual gas, causing the gas molecule to either ionise or emit fluorescent light. The byproducts from the beam-gas interaction, can be collected via an external electromagnetic field (ions and electrons), or detected using a stand alone optical system (fluorescence) to provide the one dimensional distribution information of the primary beams. Depending on the background pressure level, they usually require long integration times or extra working gas being loaded. The latter will create a large pressure bump area and cause a potential degradation of the primary beams. Recent studies \cite{HASHIMOTO2004289,FUJISAWA200350,VasilisAPL2014,Vasilis_PRAB,AmirAPL2022}, have shown that the transverse profile of particle beams can be obtained non-invasively by a novel beam profile monitor using a supersonic gas jet as a screen. Using a gas jet \cite{VasilisVacuum2014} is a novel and safer way to introduce working gases where the supersonic gas jet flows across the primary beam in a small and confined area. The signal from the interaction will be significantly increased due to the increased local density but the surrounding pressure level will minimally if not affected due to the directionality of the supersonic gas propagates. Moreover, using a gas-curtain angled at 45 degrees will have the added benefit of providing a 2D profile of the beams. The thickness, uniformity and density of the gas jet curtain screen will affect the accuracy and detection efficiency of such monitors, and thus characterising these parameters is essential. In addition, supersonic gas jets are widely used in high energy physics\cite{HEApp1,HEApp2}, Nuclear physics\cite{NuclearApp1,NuclearApp2}, Nuclear astrophysics\cite{NuclearAstrophysicsApp}, and atomic physics\cite{AtomicPhysicsApp}. 

Previously, the density of the gas jet was measured overwhelmingly by laser interferometry techniques \cite{KimAPL2003interferemetry,DitmireOL98interferemetry,LandgrafRSI2011interferemetry,GaoAPL2012interferemetry} but also by other techniques such as Rayleigh scattering \cite{Stern2007JetMeasureRayleigh}, usage of a common microphone \cite{Rajeev2013jetMicrophone}, multi-photon ionization \cite{Schofield2009JetIon,Meng2015JetIon}, nuclear scattering \cite{Pronko1993JetScattering}. The target density measured using these methods was in the range of \(10^{20} - 10^{22}\) \(m^{-3}\). When dealing with the gas jet with density in the range of \(10^{14} - 10^{17}\) \(m^{-3}\), these methods suffer from the signal to noise ratio. Compression gauge method \cite{HASHIMOTO2004289,FUJISAWA200350,Y.Hashimoto2013} was used to measure the density of a pulsed gas jet in the range of ~\(10^{16}\) \(m^{-3}\) for a similar gas jet beam profile monitor. In this paper, we extend this method to measure the absolute density and distribution of a continuous gas jet. This enables us to understand the beam profile measured by such gas jet monitor and mitigate any distortion due to the jet thickness and non-uniformity of the gas jet. The paper is structured as follow. In section \ref{sec:setup}, we describe the experimental setup. In sections. \ref{sec:gasjetform} and \ref{sec:Mprinciple}, the principle of the gas jet forming and measuring the gas jet density using the compression gauge will be explained. In section \ref{sec:Results}, the experimental results are presented and discussed together with a comparison from theoretical predication. The conclusions are then summarized in section \ref{sec:summary}.

\section{Experimental Setup}
\label{sec:setup}
\subsection{Supersonic gas jet monitor system description}
\begin{figure}[ht!]
    \centering
    \includegraphics[width=13cm]{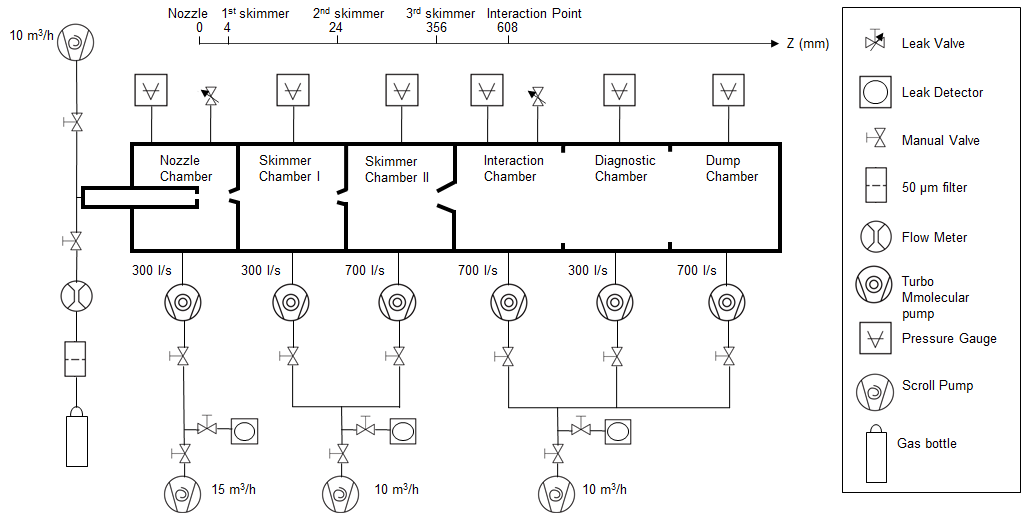}
    \caption{The layout of the gas curtain beam profile monitor setup, including the vacuum pumping system.}
    \label{fig:Vacuumsy}
\end{figure}

The layout with emphasis on pumping details of the setup are shown in Fig. \ref{fig:Vacuumsy} and similar setup were described previously \cite{VasilisVacuum2014,Vasilis_PRAB}. There are three sections including jet generation section, interaction section and gas dump section. The nozzle skimmer assembly separate the jet generation section into three chambers as nozzle chamber, skimmer chamber I and skimmer chamber II. The supersonic gas jet is generated by injecting high pressure gas (1 - 10 bar) from a gas tank through a small nozzle with a diameter of 30 \(\mu\)m into a low pressure nozzle chamber (~\(10^{-3}\) mbar). With a collimation by two conical skimmers with diameters of 180 \(\mu\)m and 400 \(\mu\)m and a pyramid-shaped skimmer with the tip size of 0.4 \(\times\) 4 \(mm^2\), the gas jet can travel mono-directionally and be shaped into a screen-like curtain for diagnostic purposes. The differential pumping stages separated by the skimmers were designed to remove the diffused gas molecules and maintain an ultra-high vacuum environment in the interaction chamber. Dumping sections, including diagnostic chamber and dump chamber, are used for dumping the jet and characterizing the jet. The pressure in each chamber is listed in Table \ref{table:pressure} with the gas jet off or continuous on at a stagnation pressure of 5 bar. This clearly shows that the introduction of the gas jet has a negligible effect on the ultra-high vacuum condition of the interaction chamber.   

\begin{table}[ht!]
\centering
\caption{ Pressure (mbar) in each vacuum chamber, with gas jet off and on at a stagnation pressure of 5 bar.}
\begin{tabular}{cccccc}
\hline 
 & Nozzle & Skimmer I & Skimmer II & Interaction & Dump \\ \hline \\
Off                                                         &   5.0$\times10^{-8}$      &      5.0$\times10^{-8}$      & 4.0$\times10^{-8}$           &      \textless{}1.0$\times10^{-9}$       &   \textless{}1.0$\times10^{-9}$    \\
                                                              &        &           &            &             &      \\
On                                                           & 3.9$\times10^{-3}$   & 8.4$\times10^{-6}$    & 7.3$\times10^{-7}$      & 4.0$\times10^{-9}$            & 1.4$\times10^{-9}$      \\ 
\hline 
\end{tabular}
\label{table:pressure}
\end{table}

\begin{figure}[ht!]
    \centering
    \includegraphics[width=12cm]{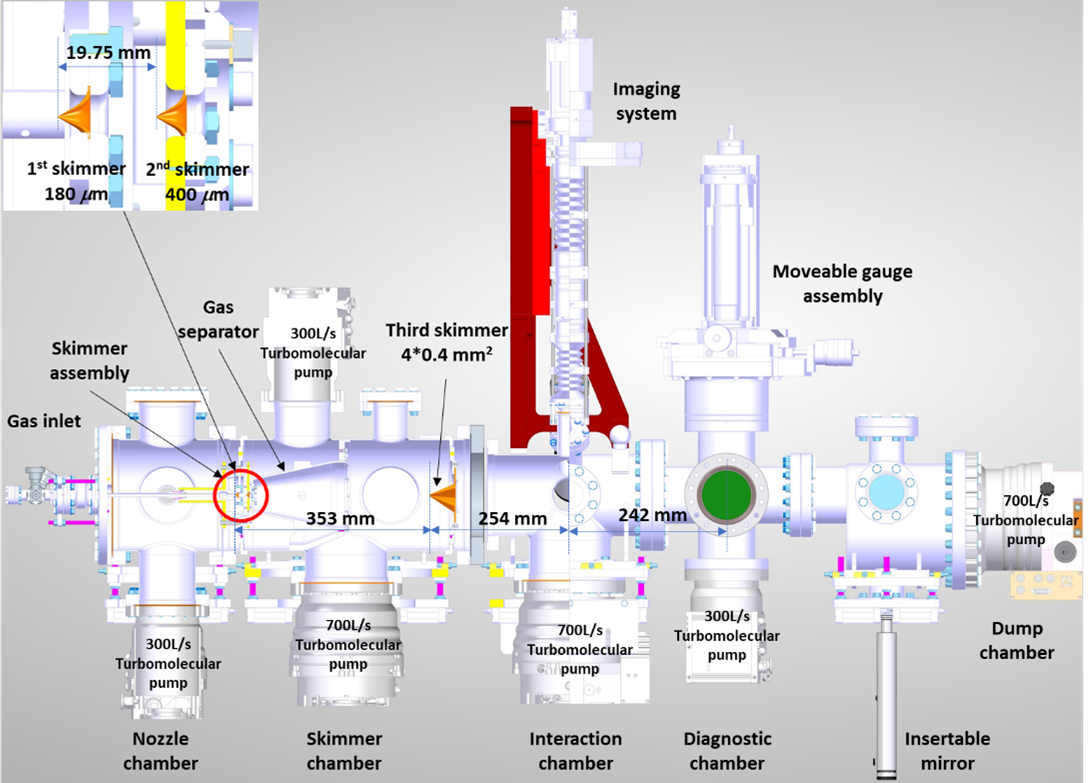}
    \caption{Schematic drawing of the gas curtain beam profile monitor system. }
    \label{fig:setup}
\end{figure}

\begin{figure}[ht!]
    \centering
    \includegraphics[width=8cm]{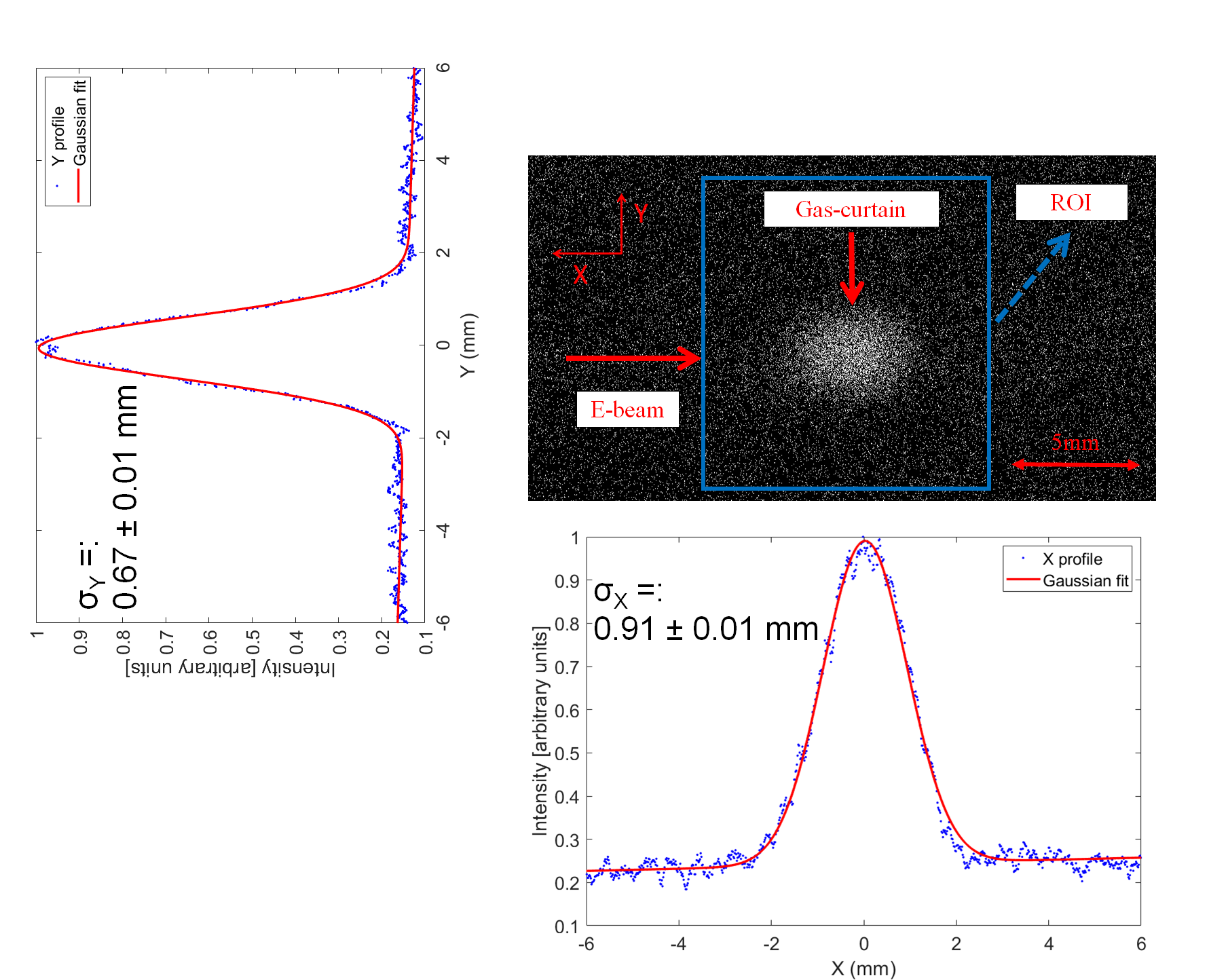}
    \caption{Image of the nitrogen gas-jet based BIF monitor with an electron beam of 5 keV and 0.6 mA. The integration time is 400 s and the inlet pressure is 5 bar.}
    \label{fig:beamimage}
\end{figure}

In the interaction section, the fluorescence induced by the electron beam interacting with the gas molecules was observed by the imaging system including a dedicated band-pass filter and intensified camera. The schematic drawing of the setup can be seen in Fig. \ref{fig:setup}. One fluorescence image of a 5 keV and 0.73 mA electron beam obtained by using a nitrogen gas jet with a stagnation pressure of 5 bar in the fluorescent wavelength of 391.4 nm is shown in Fig. \ref{fig:beamimage}. The root mean square (RMS) beam size is measured as 0.91 mm and 0.67 mm for x and y,  respectively.

\subsection{Scanning gauge system}
\begin{figure}[ht!]
    \centering
    \includegraphics[width=12cm]{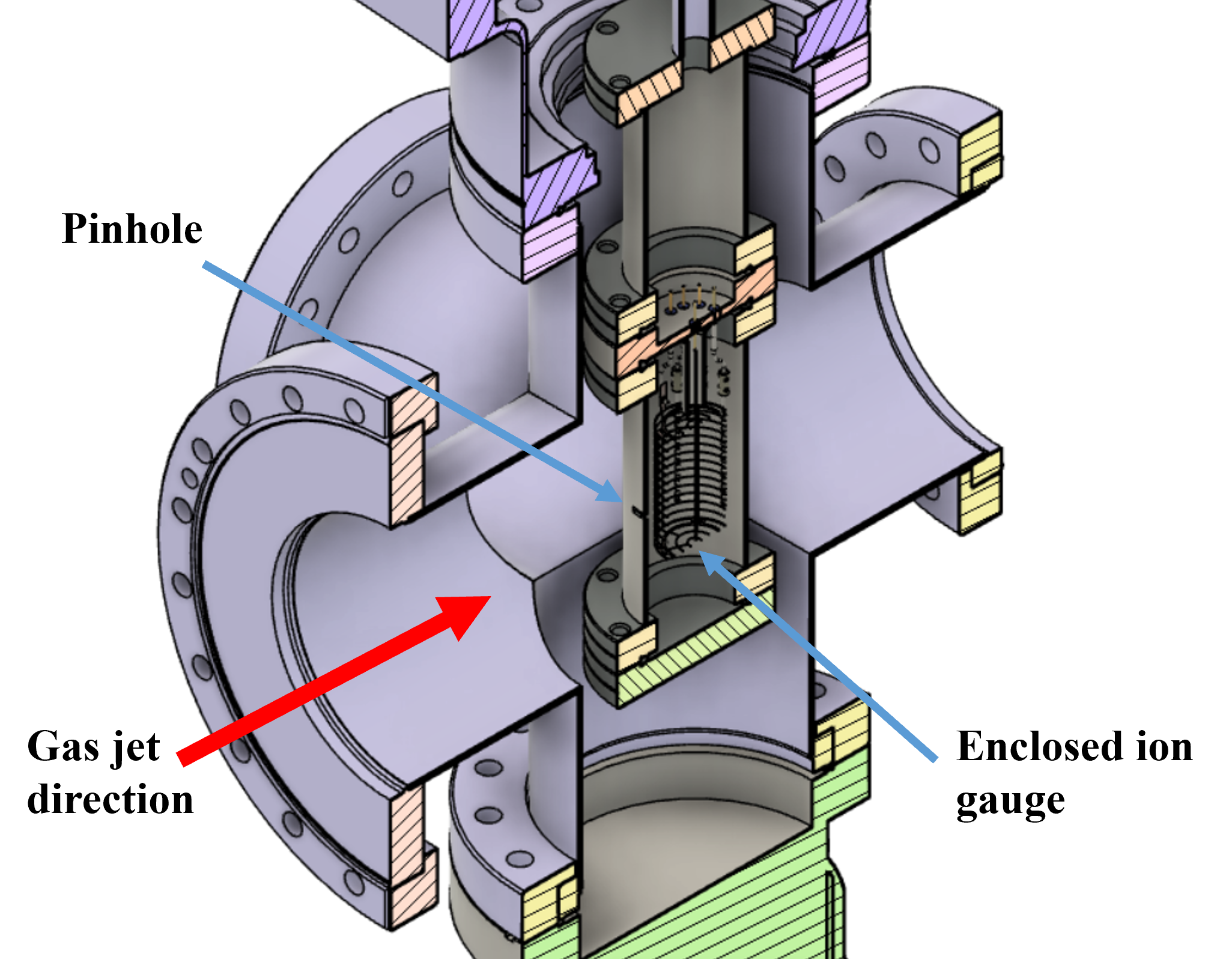}
    \caption{Schematic drawing of the movable gauge.}
    \label{fig:gauge}
\end{figure}

As indicated in Fig. \ref{fig:setup}, a movable gauge assembly could be installed either after the interaction chamber or between the second and third skimmer. As seen in Fig. \ref{fig:setup} and \ref{fig:gauge}, the assembly includes a small chamber consisting of a DN40CF straight connector with the bottom side closed by a fixed flange and the top side attached to a Bayard-Alpert (BA) type ionization gauge. The connector has a length of 125.2 mm and an inner diameter of 34.9 mm. Two BA gauges were used, one is a series 274 gauge from Granville Phillips Ltd. for the dump section and the other one is a AIG18G gauge from Arun Microelectronics Ltd. for the skimmer chamber. This gauge assembly is then connected to a VACGEN Miniax XYZ manipulator powered by three stepper-motors to allow 3-dimensional movement with a minimum resolution of 5 \(\mu\)m. On the tube of the connector, 40 mm above the bottom, there is a pinhole with a diameter of 0.5 mm. The gauge for the dump section is powered by an IGC26 ion gauge controller with a emission current of 0.1 mA by Vacgen Ltd., while the one for the skimmer chamber is powered by a NGC2 ion gauge controller with an emission current of 0.5 mA by Arun microelectronics Ltd. Their signals are amplified by a pico-ampere meter, CP8 by Cooknell Electronics Ltd. and then recorded by a oscilloscope, DS1074 Z-plus by Rigol Ltd..

\section{Forming the gas jet}
\label{sec:gasjetform}
As mentioned in Section \ref{sec:setup}, the supersonic gas jet is generated when a high-pressure gas expands through a 30 \(\mu\)m nozzle into a low-pressure region. From the nozzle till the first skimmer, the centre-line density scales as the distance increases, which can be described by Eq. (\ref{eq:1}) with the assumption of isentropic flow, ideal gas behaviour, constant heat capacity and continuum flow \cite{TheBook}.
\begin{equation}
    \rho = \frac{P_0}{k_BT_0}(1+\frac{\gamma-1}{2}M^2)^{-\frac{1}{\gamma-1}}\label{eq:1}
\end{equation}
where \(\gamma\) is the heat capacity ratio, \(\rho\) is the number density, \(P_0\) and \(T_0\) are pressure and temperature at nozzle throat, and \(M\) is the Mach number which can be calculated as follow \cite{TheBook}: 
\begin{equation}
    M = A(\frac{x-x_0}{d})^{\gamma-1}-\frac{\frac{1}{2}(\frac{\gamma+1}{\gamma-1})}{A(\frac{x-x_0}{d})^{\gamma-1}} \qquad (\frac{x}{d})>(\frac{x}{d})_{min}\label{eq:2}
\end{equation}
Here \(d\) is the nozzle throat size, \(A\) and \(x_0\) are fitted parameters which are \(\gamma\)-dependent as seen in Table \ref{table:1}.
\begin{table}[ht!]
\centering
\caption{Parameters for centre-line Mach number calculation for axisymmetric flow\cite{TheBook}.}
\begin{tabular}{cccc}
\\
\hline
\(\gamma\) & \((x_0/d)\) & \(A\) & \((x_0/d))_{min}\) \\ \hline
1.67       & 0.075             & 3.26  & 2.5                      \\
1.40       & 0.40              & 3.65  & 6                        \\ \hline
\end{tabular}
\label{table:1}
\end{table}

The temperature and the velocity of the gas jet in the continuum flow region can be obtained from Eqs.(\ref{eq:3}) and (\ref{eq:4}) \cite{TheBook}.
\begin{equation}
    T = T_0(1+\frac{\gamma-1}{2}M^2)^{-1}\label{eq:3}
\end{equation}
\begin{equation}
    v_{jet}=\sqrt{\frac{2\gamma}{\gamma-1}\frac{k_BT_0}{m}}\label{eq:4}
\end{equation}
Where \(T_0\) is the temperature of the nozzle, \(k_B\) is the Boltzmann constant and \(m\) is the mass of the gas molecule. After the 1st skimmer, the pressure decreases below \(10^{-5}\) mbar, i.e. the mean free path will be around one meter. The collisions between molecules can then be ignored, and the gas jet flow can be regarded as a molecular flow. As a result, geometric expansion can be assumed. The final density \(\rho\) will decrease from the initial value \(\rho_{skimmer}\) leaving the first skimmer with the transverse velocity spread \(\overline{v}/v_{jet}\) as a function of the  distance from the skimmer \(x\), according to Eqs. (\ref{eq:5}) and (\ref{eq:6}) \cite{GRUBER_GSI_1989}. 
\begin{equation}
    \rho = \rho_{skimmer}(1+\frac{x\overline{v}}{r_{skimmer}v_{jet}})^{-2}\label{eq:5}
\end{equation}
\begin{equation}
    \overline{v} = \sqrt{\frac{3k_B T_{skimmer}}{m}}\label{eq:6}
\end{equation}

\section{Measuring principle}
\label{sec:Mprinciple}
The small movable gauge assembly is placed in front of the gas jet, as it passes through the system. A fraction of the jet enters the small chamber through the pinhole, and then the gas molecules within the fraction are accumulated inside the small chamber. It results in a higher density and thus a higher ion gauge current reading. The new equilibrium pressure \(P\) inside the small chamber is reached when the net effusive flow through the hole is equal to the entering fraction of the jet. According to \cite{TheBook}, this pressure can be expressed as 
\begin{equation}
    P = \frac{4QIk_BT_1}{<v>A}\label{eq:7}
\end{equation}
where \(Q\) is a factor related to the shape of the pinhole. For a round channel with radius r and length L, it is defined as 
\begin{equation}
    Q = \frac{3L}{8r}\label{eq:7a}
\end{equation}
\(A\) is the area of the aperture channel and \(<v>\) is the mean velocity in the cell at the movable gauge temperature \(T_1\)   
\begin{equation}
    <v> = \sqrt{\frac{2k_BT_1}{\pi m}} \label{eq:8}
\end{equation}
\(I\) is the flux of the molecules entering through the pinhole
\begin{equation}
    I = v_{jet} \rho_{jet} A \label{eq:9}
\end{equation}
where \(v_{jet}\) and \(\rho_{jet}\) are the longitudinal velocity and the density of the jet, respectively. For the gauge, the increased pressure can be calculated from the measured ion collector current \(I_c\) using Eq. (\ref{eq:11}).
\begin{equation}
\Delta P = \frac{\Delta I_c}{S_g I_e} \label{eq:11}    
\end{equation}
where \(S_g\) is the sensitivity factor which is 1.0 for nitrogen and 0.3 for neon, and \(I_e\) is the electron emission current of the chosen gauge. Combining Eqs. (\ref{eq:4}),(\ref{eq:7})-(\ref{eq:11}), we obtain the density of the gas jet as 
\begin{equation}
    \rho_{jet} = \frac{\Delta I_c}{S_g I_e}\frac{1}{4Q k_B T_1}\sqrt{\frac{T_1}{T_0}\frac{\gamma-1}{2\gamma \pi m}} \label{eq:12}
\end{equation}

In the analysis, we made several assumptions. First that the gas jet is uniform and stable on the scale of the pinhole size. Second that the out-gassing rate for the inner surface of the small chamber does not change. Third that at each pinhole location, the new equilibrium inside the small chamber will be established on a time scale of less than one second. These assumptions are reasonable for a continuous jet and a degassed ion gauge. The molecule that enters the small chamber could experience more than 1000 interactions within the chamber surface in one second which is long enough to achieve a new equilibrium .  

For the density scan, the manipulator moves the small chamber with the pinhole to a set of coordinates, 0.25 mm per step horizontally and vertically in a region of interest across the gas jet and beyond. At each coordinate, the collector current of the ion gauge is recorded over 4 s to wait for a new equilibrium to achieve. The current is then averaged and subtracted from the ambient current and the density at that coordinate is calculated from the current difference using Eq. (\ref{eq:12}). Finally, the density map from all coordinates forms the transverse density map of the gas jet.  

Two error sources were considered, position error and density error. The gauge is mounted on a manipulator, the resolution and repeatability of the motion in each axis is 5 \(\mu\)m. For the measurement, the step size is always larger than 100 microns, i.e. we can ignore the position error. For a typical B-A type gauge, the measurement error of the current \(I_c\) can be 20\%. Because of the hot ions, the small chamber could be heated up and the temperature $T_1$ will be higher than the normal room temperature of 300 K. We didn't have a direct measurement of that temperature, and instead, a 5\% error was used for the temperature. Combining these two errors, one can get a density error about 20\% using Eq. (\ref{eq:13}).
\begin{equation}
    \frac{\sigma_{\rho_{jet}}}{\rho} = \sqrt{\frac{\sigma^2_{\Delta I_c}}{\Delta I_c^2}+\frac{\sigma^2_{T_1}}{4T^2_1}} \label{eq:13}
\end{equation}
where \(\sigma\) is the absolute error for each measurement.

\section{Results and Discussions}
\label{sec:Results}
\begin{figure}[ht!]
    \centering
    \includegraphics[scale=0.55]{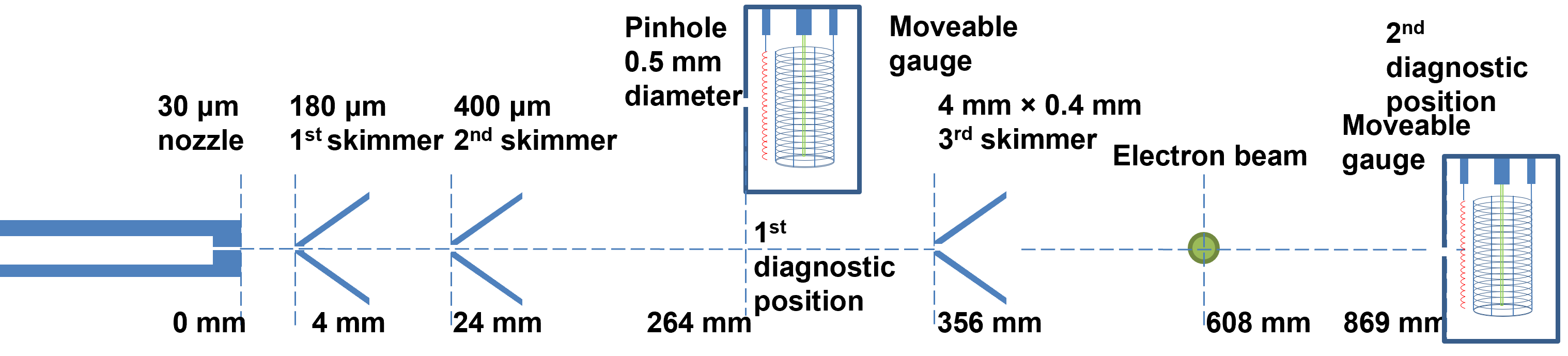}
    \caption{Schematic of the gas jet system with two movable gauges}
    \label{fig:measurement_geometry}
\end{figure}

The ideal position to measure the gas jet density distribution will be at the interaction point. Due to space limitations, we chose to measure it before and after the interaction point. Because the gas jet is in the molecular flow region after second skimmer, the gas jet density distribution can be calculated from these measurements from linear expansion. The geometry of the whole system is summarised as Fig. \ref{fig:measurement_geometry}.

\begin{figure}[ht!]
    \centering
    (a)
    \includegraphics[width=8cm]{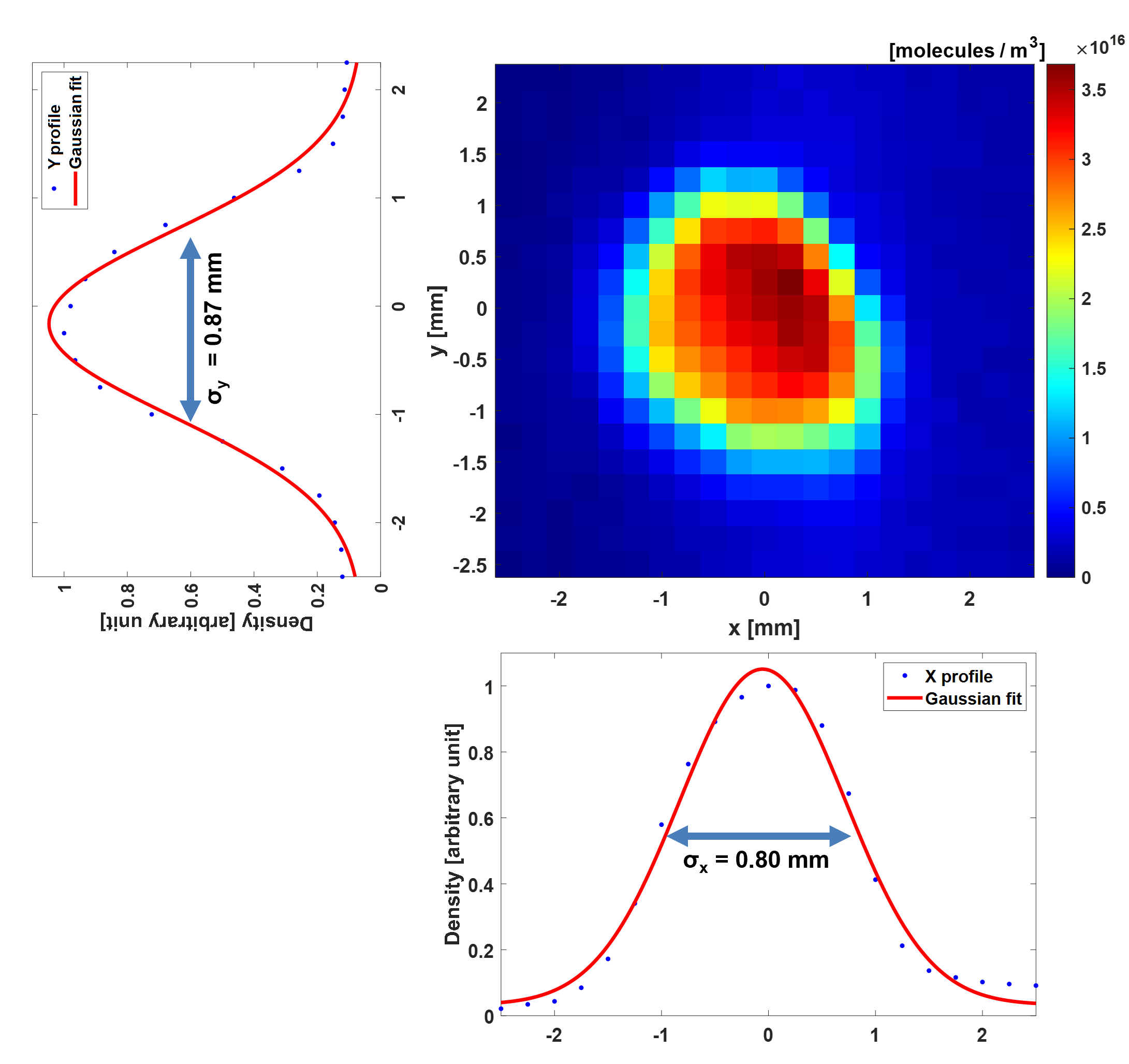}
    
    (b)
    \includegraphics[width=8cm]{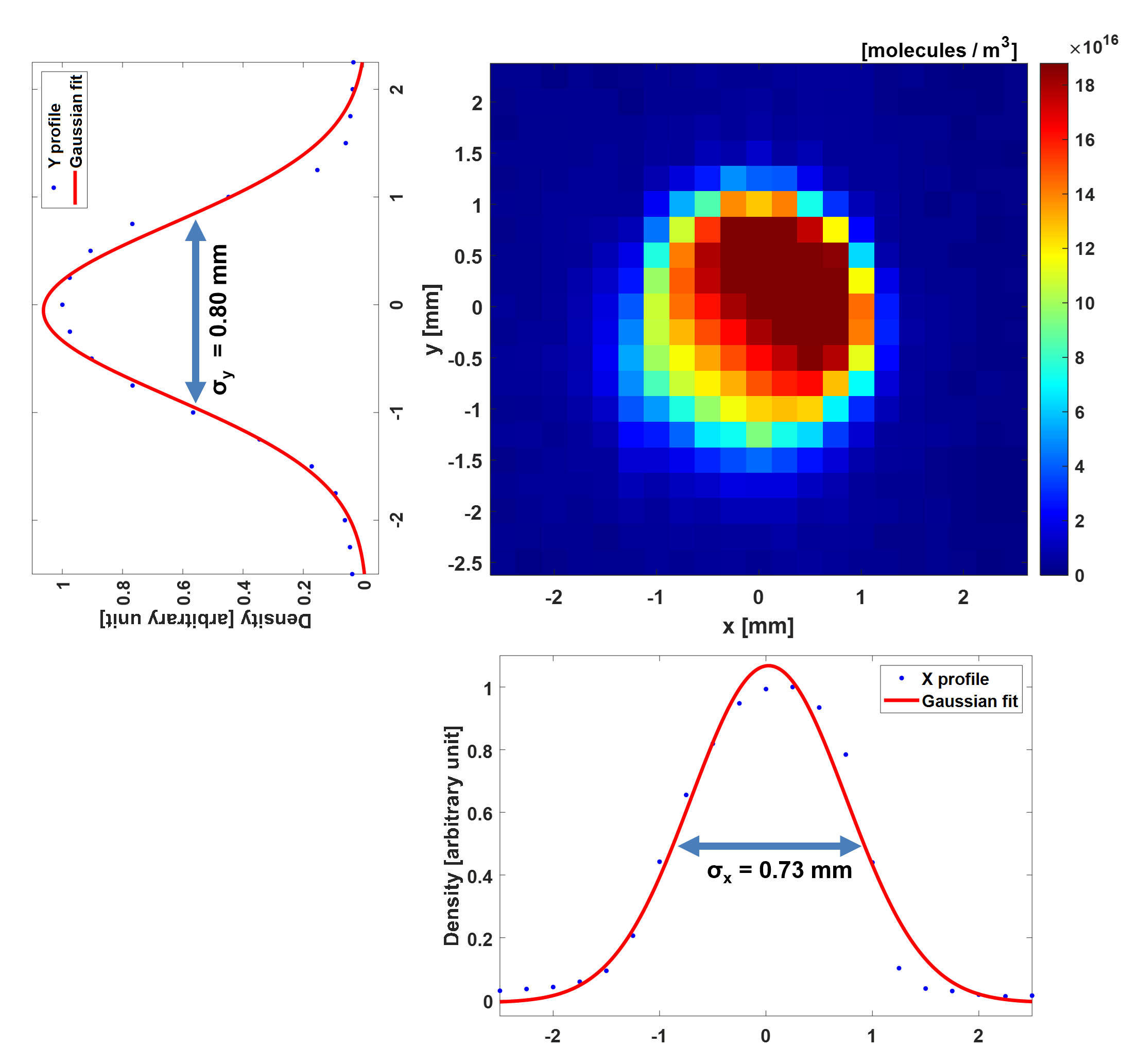}
    \caption{Measurement of nitrogen (a) and neon (b) gas jet density distribution at first diagnostic position}
    \label{fig:NitrogenNeon1st}
\end{figure}

Since the nozzle, 1st and 2nd skimmers are all circular in shape, the ideal transverse distribution of the gas jet between the 2nd skimmer and the 3rd skimmer is supposed to be rotational symmetrical. A measurement of such jets with nitrogen and neon as working gases at 1st diagnostic position are shown in Fig. \ref{fig:NitrogenNeon1st}. The inlet pressure was 5 bar in both cases and the step size of the movable gauge was 0.25 mm. The maximum densities are \(2.2\times10^{16}\) \(m^{-3}\) for nitrogen gas jet and 1.1 \(\times\) \(10^{17}\) \(m^{-3}\) for neon. The differences in density for the two gases depend on the gas molecule status, mono-atomic or diatomic, the molecular weight m, the initial pressure \(P_0\) and temperature \(T_0\). In the continuum flow region, roughly from the nozzle to the first skimmer, the density and the temperature of the gas jet drop much quicker for a diatomic gas than a mono-atomic gas, see Eq. (\ref{eq:3}). The temperature of the gas jet and the molecular weight will determine the thermal velocity, see Eq. (\ref{eq:5}), which is key factor for the jet expansion in the molecular flow region and reduce the jet density further. Both density profiles show a quasi-circular shape. The asymmetry could result from alignment errors related to the nozzle and both skimmers. A Gaussian fit for both dimensions shows a jet size of (0.80 mm, 0.87 mm) for nitrogen and (0.73 mm, 0.81 mm) for neon, respectively. The slightly smaller size of the neon gas jet is due to the smaller velocity spread of the neon gas jet as compared to nitrogen. As shown in Eq. (\ref{eq:6}), this is a combined effect from the temperature at the skimmer, the molecule mass, as well as the collimation done by the skimmers. The latter is a more complicated geometrical effect where dedicated simulations are required to calculate the differences.These are beyond the scope of this paper.      

\begin{figure}[ht!]
    \centering
    (a)
    \includegraphics[width=6cm]{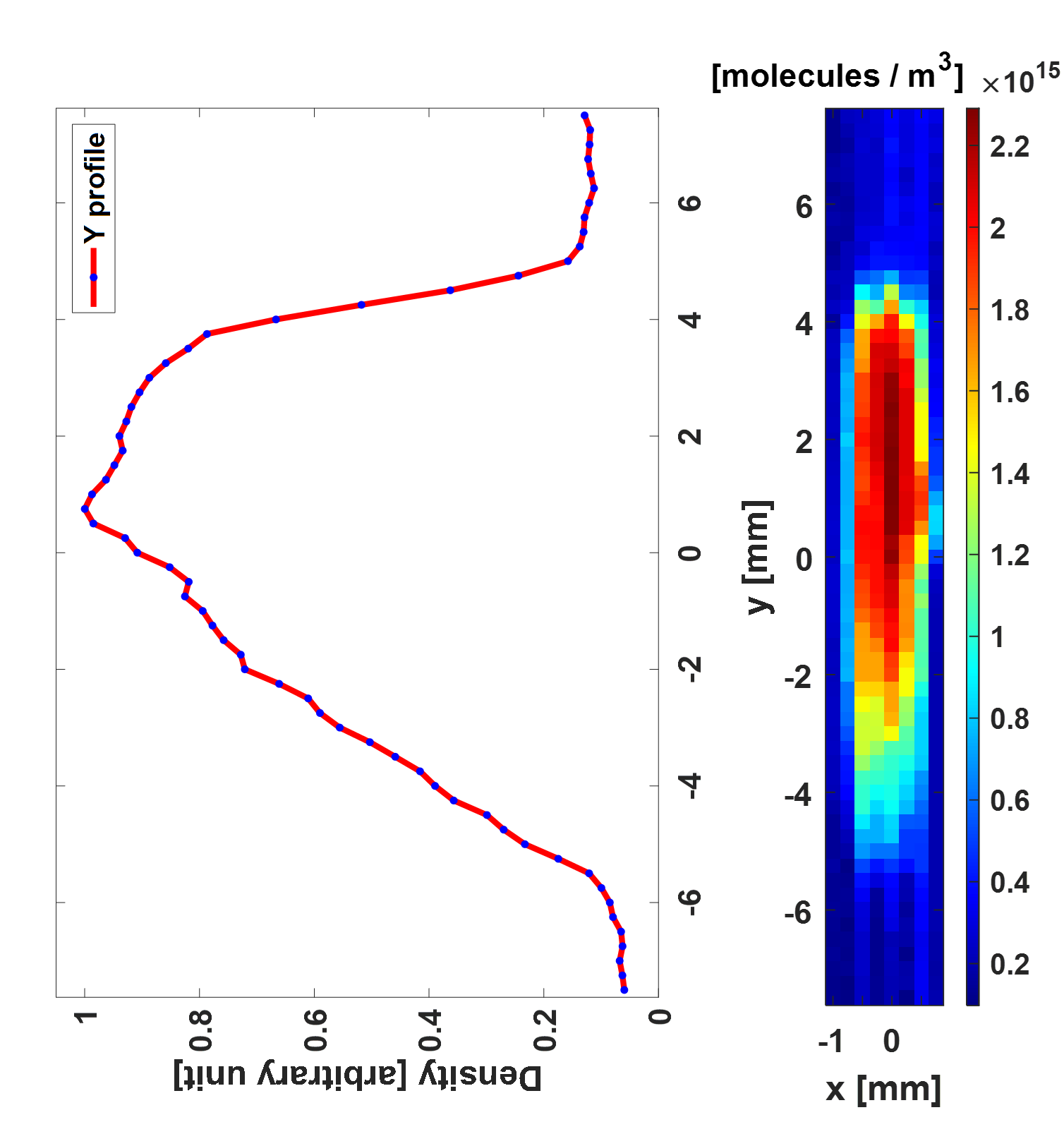}
    (b)
    \includegraphics[width=6cm]{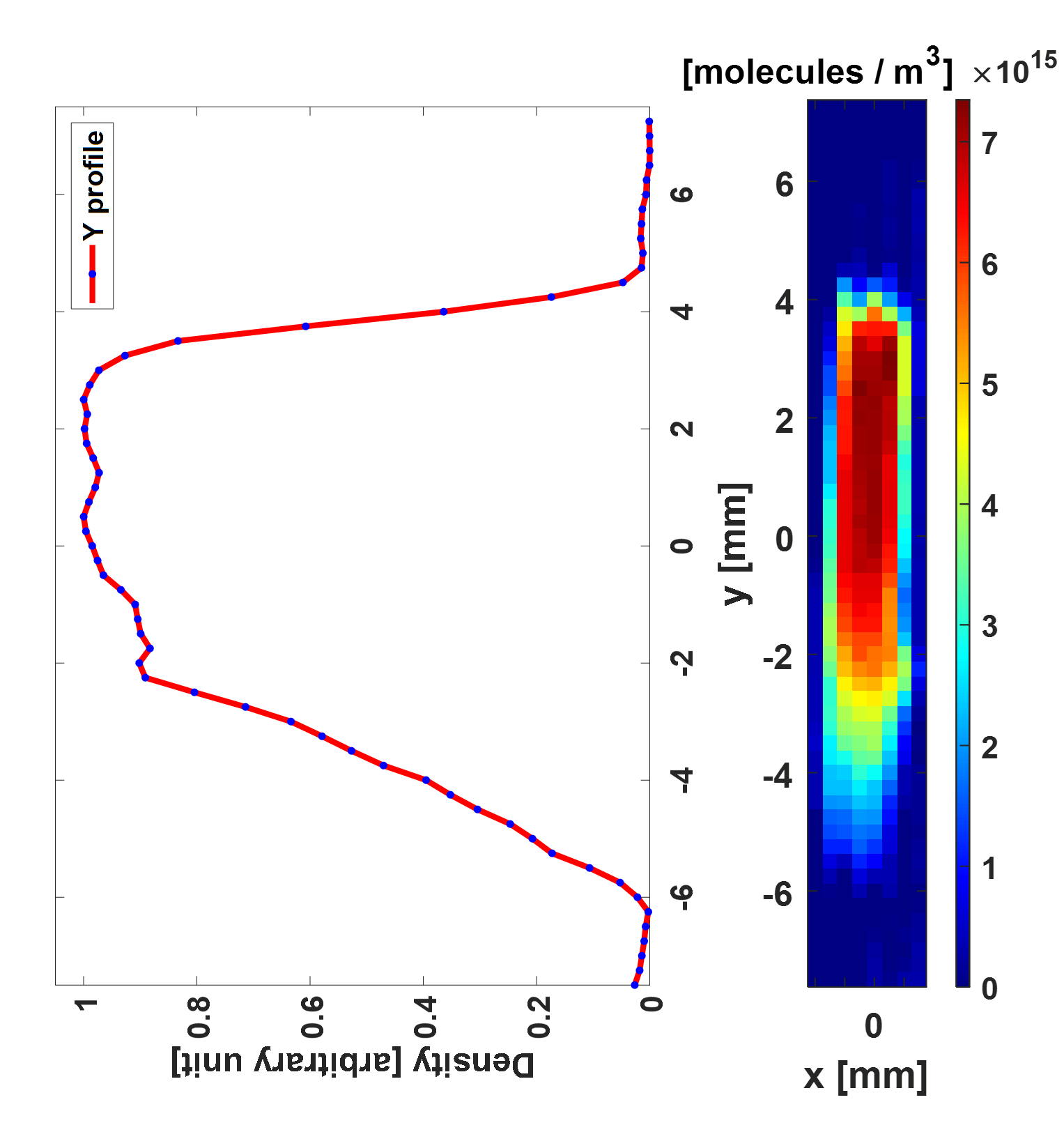}
    \caption{Measurement of nitrogen (a) and neon (b) gas jet density distribution at the second diagnostic position}
    \label{fig:NitrogenNeon2nd}
\end{figure}

\begin{figure}[ht!]
    \centering
    \includegraphics[scale=1.0]{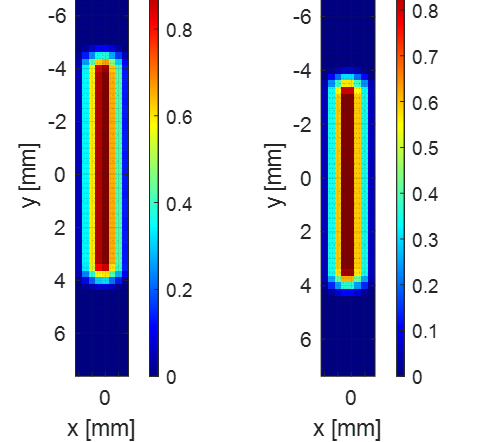}
    \caption{Convoluted images of uniform gas jet distributions with the pinhole matched with the measurement: left: Nitrogen;right: Neon. }
    \label{fig:Convolution}
\end{figure}

The density distribution measurement at the 2nd diagnostic position is shown in Fig. \ref{fig:NitrogenNeon2nd}. The quasi-rectangular shape of the gas jet is the result of a vertically placed rectangular 3rd skimmer (0.4 mm \(\times\) 4 mm). Both distributions show a sharp drop in density at the top and a Gaussian tail at the bottom, which indicates that the 3rd skimmer is off the center of the nozzle-skimmer assembly. Similarly, the density for the neon gas jet is still higher than the nitrogen gas jet which is consistent with the measurements at the 1st diagnostic position and the geometric expansion assumptions. Here, the full width at half maximum (FWHM) sizes were measured for both cases to be (7.50 \(\pm\) 0.25 mm, 1.25 \(\pm\) 0.25 mm) for nitrogen and (7.63 \(\pm\) 0.25 mm, 1.25 \(\pm\) 0.25 mm) for Neon. The size differences are in the error range which shows there is little difference for the velocity spread of both gas jets after they are collimated by the 3rd skimmer. Note that, due to the finite size of the pinhole for the movable gauge, the distribution measured here will be a convolution of the real intensity distribution and the pinhole size. De-convolution of these measurement by assuming a uniform original distribution of these measurement gives the original size of the jet is (8.32 mm. 0.96 mm) and (7.54 mm, 0.87 mm) for nitrogen and neon. The convoluted density distribution of the uniform gas jet with the pinhole is shown in Fig. \ref{fig:Convolution}, which shows a good match with the measurements in Fig. \ref{fig:NitrogenNeon2nd}.

For beam profile measurements, the vertical size of jet defines the range one can use to characterize the primary beam. The thickness and the uniformity of the gas jet will determine on smearing effects and hence the image quality. From our jet density measurement,both measurements show a good uniformity in the central area in the 2nd diagnostics position. From 3rd skimmer to the interaction point and then the 2nd diagnostics position, the jet flow is molecular flow where no collision will occur. Thus, for the interaction point, the gas jet thickness can be estimated by linear expansion from 0.4 mm at the 3rd skimmer location to 1.25 mm at the 2nd diagnostic position, which gives a thickness of 0.85 \(\pm\) 0.14 mm at the interaction point. 

\section{Conclusion}
\label{sec:summary}
In this paper, we described a method to measure the absolute density and the density distribution of a supersonic gas jet in the molecular flow region. This method was applied to a supersonic gas jet used for beam profile monitors and measured the density of both nitrogen gas jet and neon gas jet in the range from \(10^{14}\) to \(10^{17}\) \(m^{-3}\) with a good spatial resolution of sub-mm. The spatial resolution is limited by the pinhole size but can be reconstructed by the de-convolution method if a known distribution (uniform distribution here) are assumed. The measurement of gas jet density and its distribution will allow the further study or optimization of the gas jet by changing the geometry of the skimmers to meet different diagnostics requirements such as charged particle beam intensity, integration time and ambient vacuum environment. Currently, this method is used regularly for characterizing a supersonic gas jet in the beam profile monitor \cite{AmirIPAC21} designed for the LHC proton beam where a higher gas jet density and a larger and thinner gas jet curtain with uniform density distribution are required. This method could also applied for a similar supersonic gas jet used in other field such as nuclear target and atomic physics. Recent development of a beam profile and dose monitor for medical used charged particle beam \cite{kumar2020} is another example where the measurement of the gas jet density and its distribution will help in the designing phase and commissioning phase.



 \bibliographystyle{elsarticle-num} 
 \bibliography{density_measurement_paper}





\end{document}